\documentclass[conference]{IEEEtran}
\IEEEoverridecommandlockouts
\usepackage{cite}
\usepackage{amsmath,amssymb,amsfonts}
\usepackage{algorithmic}
\usepackage{graphicx}
\usepackage{textcomp}
\usepackage{xcolor}
\def\BibTeX{{\rm B\kern-.05em{\sc i\kern-.025em b}\kern-.08em
    T\kern-.1667em\lower.7ex\hbox{E}\kern-.125emX}}
\usepackage{subcaption}
\usepackage{float} 
\usepackage{algorithmic}
\usepackage{amsmath}
\usepackage{hyperref} 
\usepackage{booktabs} % For formal tables

\usepackage{tikz}
\usepackage[framemethod=tikz]{mdframed} % add the frame for the figure
\usetikzlibrary{shapes.geometric, arrows.meta, positioning, calc, shadows}
\begin{document}

\title{BirdsongChat: A Hybrid Multi-Agent Framework for Multimodal Embodied Behavior Simulation}

% \title{Conference Paper Title*\\

% \and
% \and

\author{\IEEEauthorblockN{Callie C. Liao$^{*}$}
\IEEEauthorblockA{\textit{Dept. of Computer Science} \\
\textit{Stanford University}\\
Stanford, CA, USA \\
ccliao@cs.stanford.edu}
\and
\IEEEauthorblockN{Duoduo Liao$^{*}$}
\IEEEauthorblockA{\textit{School of Computing} \\
\textit{George Mason University}\\
Fairfax, VA, USA \\
dliao2@gmu.edu}
\and
\IEEEauthorblockN{Ellie L. Zhang$^{*}$}
\IEEEauthorblockA{ 
\textit{IntelliSky}\\
McLean, VA, USA}
}

\maketitle

\def\thefootnote{*}\footnotetext{Equal contribution.}
\def\thefootnote{†}\footnotetext{The \href{https://www.intellisky.org/share/birdsongchat_demo.mp4}{demo video} is available via the provided link.}

\begin{abstract}
Multimodal embodied systems require translating human intentions into interpretable and coordinated behaviors across heterogeneous modalities. However, existing multimodal agents often rely on implicit representations, limiting controllability and cross-modal consistency. We present a hybrid multi-agent framework for interactive multimodal behavior simulation that bridges semantic reasoning and physical execution through a Unified Parameter Representation (UPR). LLM-based reasoning agents transform multimodal inputs into UPR, which encodes behavioral states and interpretable control parameters for simulation agents generating synchronized 3D motion, spatialized soundscapes, and environmental behaviors. We develop \emph{BirdsongChat}$^{\dagger}$ as a prototype implementation of the proposed framework, using interactive avian behavior simulation as a testbed that tightly couples motion, vocalization, and environmental context. BirdsongChat is evaluated on text- and image-guided scenarios involving species, behaviors, affective states, environments, and multi-bird interactions. The system achieves normalized scores of 94.4\% for cross-modal coherence, 100\% for affective consistency, and 92.6\% for generation consistency. These results demonstrate that an explicit intermediate representation effectively bridges semantic reasoning and physical execution, improving controllability and multimodal synchronization. The proposed framework thus offers a generalizable design principle for embodied AI systems requiring interpretable semantic-to-physical coordination across modalities, with potential applications in bio-inspired ecoacoustics, swarm robotics, virtual environments, and creative multimedia.
\end{abstract}

\begin{IEEEkeywords}
Embodied AI, Multi-Agent Systems, Large Language Models, Semantic-to-Physical Transformation, Unified Parameter Representation, Cross-Modal Consistency, Avian Behavior Simulation, Birdsong Soundscapes.
\end{IEEEkeywords}

%%%%%%%%%%%%%%%%%%%%%%%%%%%%%%%%%%%%%%%%%%%%%%%%%%%%%
%============================
% Teaser Figure
%============================
\begin{figure}
\centering
\begin{minipage}{\columnwidth}
\begin{mdframed}[linecolor=gray, linewidth=0.5pt, roundcorner=10pt]
\includegraphics[width=\textwidth]{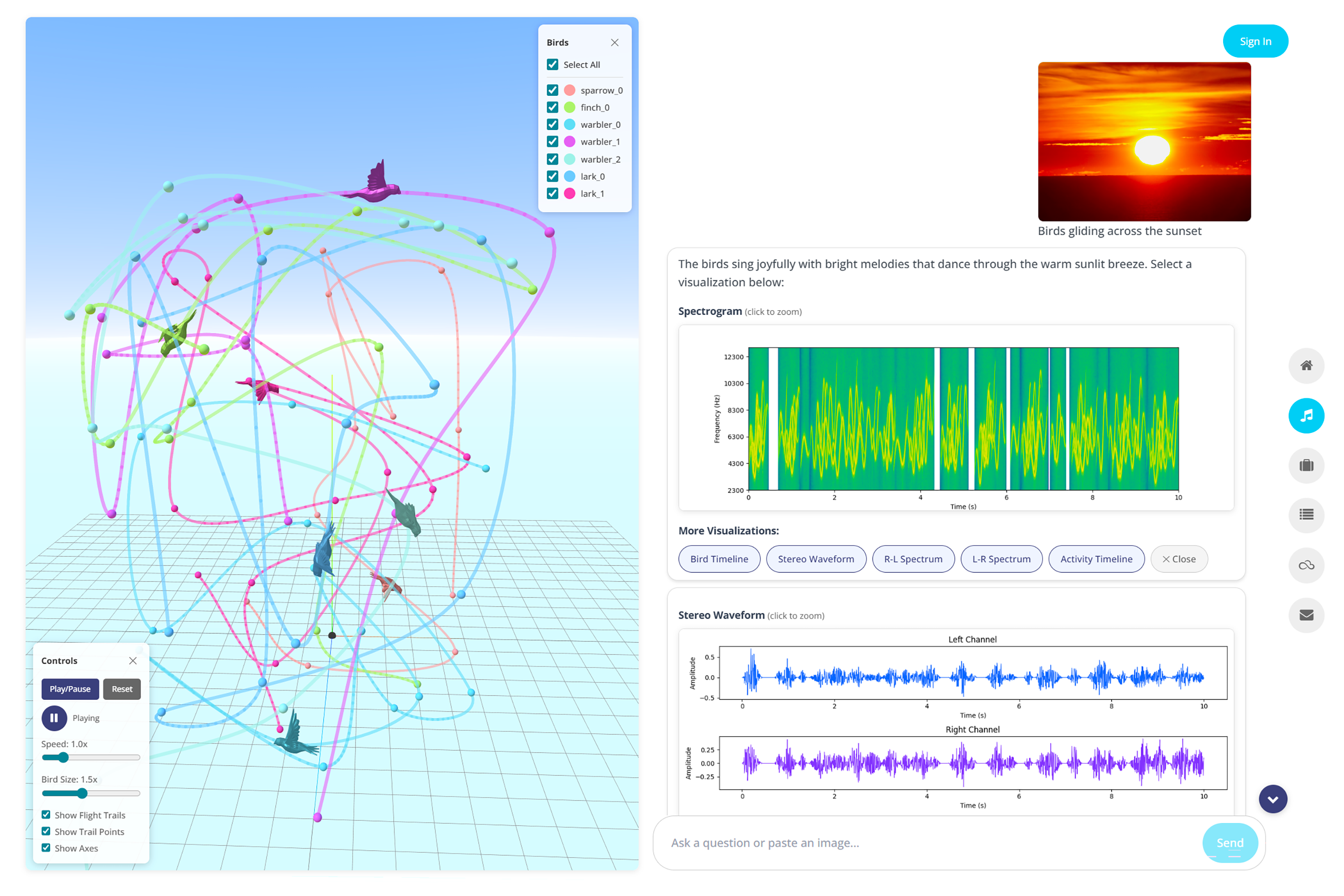}
\end{mdframed}
\end{minipage}
 \caption{
  \emph{BirdsongChat: Interactive Multimodal Control and Generation of Synchronized 3D Bird Motion and Soundscape}. 
Foundation models perform semantic interpretation of species, behavior, and context from text and images. The resulting UPR parameters control algorithmic simulators for motion, sound, and environment generation.}
  \label{fig:teaser}
\end{figure}
%%%%%%%%%%%%%%%%%%%%%%%%%%%%%%%%%%%%%%%%%%%%%%%%%%%%%

%%%%%%%%%%%%%%%%%%%%%%%%%%%%%%%%%%%%%%%%%%%%%%%%%%%%%
%----------------------------------------------------
% Figure: Framework
%----------------------------------------------------
\begin{figure*}[h]
 \centerline{
 \includegraphics[width=\linewidth]{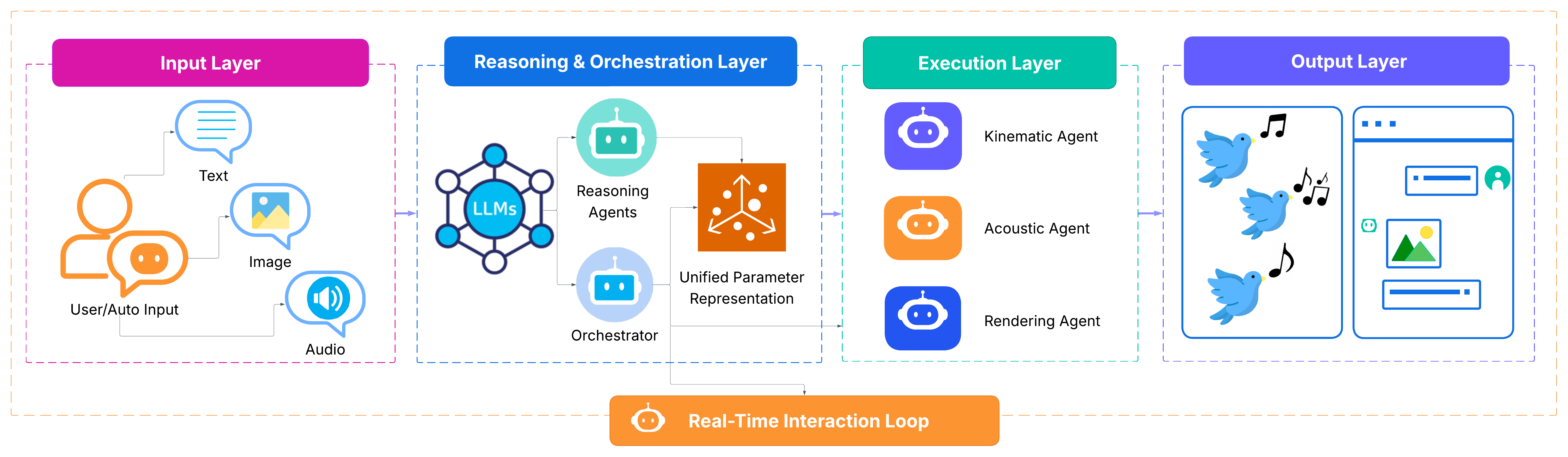}}
  \caption{ \emph{Overview of the BirdsongChat framework for physically grounded multimodal generation and control}. \textnormal{The architecture bridges high-level semantic reasoning with low-level physical execution through a hierarchical design including four functional layers, maintaining interpretability, physical plausibility, and cross-modal coherence.} 
  }
 \label{fig:sys_flow}
\end{figure*}
%%%%%%%%%%%%%%%%%%%%%%%%%%%%%%%%%%%%%%%%%%%%%%%%%%%%%

\section{Introduction}

Recent advances in Large Language Models (LLMs) and multimodal foundation models have significantly improved the ability of AI systems to interpret human instructions, reason over complex contexts, and interact with multimodal information. Vision-language models such as Flamingo~\cite{NEURIPS2022_960a172b}, embodied models such as PaLM-E~\cite{pmlr-v202-driess23a}, and unified multimodal models such as VILA-U~\cite{ICLR2025_e9e140df} demonstrate progress toward integrated multimodal understanding and generation. Recent generative models further achieve high-quality synthesis across image, audio, and motion domains~\cite{10.1145/3658146,NEURIPS2023_94b472a1}. These advances motivate embodied systems that translate multimodal intentions into coordinated physical behaviors.

However, a fundamental challenge remains: \emph{how can multimodal reasoning be connected to interpretable, physically grounded, and cross-modally consistent behavior generation?} Existing embodied agents primarily focus on planning and action execution~\cite{pmlr-v202-driess23a,pmlr-v270-kim25c}, while multimodal generative models often rely on implicit representations for end-to-end synthesis~\cite{10.1145/3658146,NEURIPS2023_94b472a1}. Although effective for understanding and generation, these approaches provide limited control over intermediate behavioral states and physical constraints, motivating explicit representations that bridge semantic reasoning and physical execution.

In this work, we study how multimodal semantic intentions can be transformed into coordinated physical behaviors in embodied environments using interactive avian behavior simulation as a concrete testbed in which motion, vocalization, and environmental context are tightly coupled. Building on agent-based behavioral models and procedural simulation approaches~\cite{10.1145/37401.37406}, we present a hybrid multi-agent framework that connects LLM-based reasoning agents and physically-based simulation agents through an interpretable Unified Parameter Representation (UPR).

The proposed framework combines reasoning agents for multimodal interpretation, orchestration, and semantic-to-physical transformation with simulation agents that execute UPR parameters to generate coordinated motion, vocalizations, spatialized soundscapes, and environment-aware scenes. The UPR serves as a shared interface between semantic representations and physical controls, preserving consistency across modalities. Unlike latent-based end-to-end generation, this architecture separates reasoning from execution, providing controllable and interpretable multimodal behavior generation. As a prototype implementation, \emph{BirdsongChat} is evaluated on text- and image-guided scenarios involving species, behaviors, affective states, environments, and multi-bird interactions. Results demonstrate that an explicit intermediate representation effectively bridges semantic reasoning and physical execution, improving controllability and multimodal synchronization. The framework can further accommodate additional modalities, with broad applicability to bio-inspired ecoacoustics, swarm robotics, virtual environments, and interactive multimedia.

The main contributions are as follows:
\begin{itemize}
\item \emph{Unified Parameter Representation (UPR)}: An interpretable intermediate representation that bridges high-level semantic reasoning and low-level physical execution through shared behavioral states and explicit control parameters, maintaining cross-modal consistency and controllability.

\item \emph{UPR-driven hybrid multi-agent framework}: A hybrid architecture that translates multimodal semantic reasoning into physical behaviors by encoding behavioral information into UPR to drive coordinated simulation agents, providing synchronized motion, sound, and visual scene generation and a generalizable design principle for embodied AI.

\item \emph{BirdsongChat}: An interactive avian simulation that demonstrates the proposed framework by generating synchronized avian motion, vocalizations, spatialized soundscapes, and environment-aware scenes from text and image inputs.
\end{itemize}

\section{Related Work}

\subsection{LLM-based Embodied Agents and Grounding}
LLMs have increasingly been integrated into embodied systems as high-level reasoning and decision-making components. SayCan~\cite{pmlr-v205-ichter23a} combines language reasoning with robot skill selection, while PaLM-E~\cite{pmlr-v202-driess23a}, RT-2~\cite{pmlr-v229-zitkovich23a}, and vision-language-action models such as OpenVLA~\cite{pmlr-v270-kim25c} extend multimodal reasoning toward embodied control. Recent foundation-model-based systems further explore reasoning, orchestration, and interaction in complex environments~\cite{10943341,NEURIPS2024_cebbd24f}. These approaches demonstrate the potential of foundation models for instruction understanding and physical action coordination. However, they primarily focus on task planning, skill invocation, or action prediction, with limited explicit representations connecting semantic intent to continuous physical behaviors. The proposed framework extends this direction by introducing the UPR that provides an interpretable interface between semantic reasoning and physically executable simulation.

\subsection{Multimodal Foundation Models and Cross-Modal Generation}
Multimodal foundation models have advanced cross-modal understanding and generation across language, vision, and audio. Flamingo~\cite{NEURIPS2022_960a172b} and BLIP-2~\cite{pmlr-v202-li23q} demonstrate strong vision-language alignment, while unified models such as VILA-U~\cite{ICLR2025_e9e140df}, OmniFlow~\cite{11094075}, and MIO~\cite{wang-etal-2025-mio} explore joint multimodal understanding and generation. Recent generative models also achieve high-quality synthesis across modalities, including text-to-image, audio, and motion generation~\cite{10.1145/3658146,10.1609/aaai.v39i24.34750,agostinelli2023musiclmgeneratingmusictext}. However, these models typically rely on implicit representations optimized for end-to-end generation, providing limited control over intermediate behavioral states and physically constrained processes. Our framework addresses this limitation through an explicit semantic-to-physical interface that maps multimodal reasoning outputs to interpretable kinematic, acoustic, and environmental parameters.

\subsection{Physically Grounded Simulation and Multi-Agent Systems}
Physically-based simulation provides interpretable mechanisms for generating complex behaviors under explicit constraints. Agent-based models such as Boids~\cite{10.1145/37401.37406} demonstrate emergent collective behavior through local interaction rules, while stochastic dynamical systems~\cite{UhlenbeckOrnstein30,KloedenPlaten92} provide foundations for controllable motion variation. Recent work on embodied intelligence and spatial understanding further emphasizes the importance of grounding agents in physical environments~\cite{10.1145/3717059}. Meanwhile, language-driven agent frameworks such as ReAct~\cite{yao2023reactsynergizingreasoningacting}, Voyager~\cite{wang2023voyageropenendedembodiedagent}, Generative Agents~\cite{10.1145/3586183.3606763}, and AgentMaster~\cite{liao-etal-2025-agentmaster} demonstrate the ability of LLMs to support reasoning, planning, memory, exploration, and multi-agent interaction. However, these lines of work largely emphasize either language-driven decision-making or physically grounded behavior generation, with limited focus on tightly coupling LLM reasoning with interpretable, physically based multi-agent simulation. Our framework bridges these two paradigms by integrating LLM-based reasoning agents with physically based simulation agents through the UPR, enabling synchronized and interpretable multimodal behavior generation.

\section{Methodology}

\subsection{System Overview}

The proposed hybrid multi-agent framework integrates LLM-based reasoning agents with physically-based simulation agents for interactive multimodal behavior simulation. It generates coordinated 3D motion, procedural spatialized soundscapes, and environment-aware scenes from text and image inputs. As shown in Figure~\ref{fig:sys_flow}, They architecture consists of four layers: (1) an \emph{Input Layer} for multimodal interaction, (2) a \emph{Reasoning and Orchestration Layer} with LLM-based agents for semantic interpretation and semantic-to-physical transformation, (3) an \emph{Execution Layer} with specialized agents for physically-based simulation, and (4) an \emph{Output Layer} for synchronized audiovisual rendering.

The key design principle is to separate semantic reasoning from modality-specific simulation. LLM-based reasoning agents transform multimodal inputs into semantic and behavioral representations, which are mapped to interpretable control parameters through the UPR. The UPR serves as an intermediate representation between reasoning and execution, encoding shared behavioral states that coordinate hybrid simulation agents. Motion, acoustic, and rendering agents derive modality-specific parameters from the same UPR state, maintaining temporal alignment and behavioral consistency across generated modalities. Rather than directly synthesizing audiovisual outputs with neural generative models, the system uses UPR-driven algorithmic simulators to produce controllable and physically grounded multimodal behaviors.

\subsection{Unified Parameter Representation (UPR)}

\subsubsection{UPR Definition}

The Unified Parameter Representation (UPR) is an intermediate representation that bridges LLM-based reasoning agents and physically-based simulation agents. It encodes shared semantic, behavioral, kinematic, acoustic, and environmental states required for coordinated multimodal behavior generation:
\begin{equation}
\mathcal{P}
=
\{
\mathcal{P}_{sem},
\mathcal{P}_{beh},
\mathcal{P}_{kin},
\mathcal{P}_{acous},
\mathcal{P}_{env}
\}.
\end{equation}

Here, $\mathcal{P}_{sem}$ represents semantic attributes inferred from multimodal inputs, $\mathcal{P}_{beh}$ represents intermediate behavioral states, $\mathcal{P}_{kin}$ contains kinematic control parameters, $\mathcal{P}_{acous}$ contains acoustic synthesis and spatialization parameters, and $\mathcal{P}_{env}$ represents environmental conditions.

The UPR defines a hierarchical semantic-to-physical transformation:
\begin{equation}
(\mathcal{P}_{sem},\mathcal{P}_{env})
\rightarrow
\mathcal{P}_{beh}
\rightarrow
(\mathcal{P}_{kin},\mathcal{P}_{acous}),
\end{equation}
where high-level multimodal descriptions are transformed into interpretable behavioral states and modality-specific control parameters. By maintaining a shared behavioral representation, the UPR preserves consistency between user intent, simulated motion, acoustic events, and environmental rendering.

For instance, given the prompt \emph{``An alarmed robin calls out from a wind-swayed tree branch,''} the UPR maps semantic intent ($\mathcal{P}_{sem}$) and environmental conditions ($\mathcal{P}_{env}$) into a shared high-alert behavioral state ($\mathcal{P}_{beh}$). This intermediate state then jointly drives physical execution of both modalities, producing rapid head-turn trajectories ($\mathcal{P}_{kin}$) alongside high-pitched, spatialized alarm calls ($\mathcal{P}_{acous}$). 
A unified behavioral state ensures cross-modal consistency across motion, acoustics, and environmental responsiveness.

Unlike task-oriented representations that primarily encode action sequences or symbolic plans, UPR encodes semantic attributes, behavioral states, and continuous physical controls across heterogeneous modalities. Rather than merely specifying behaviors, UPR provides a shared representation for coordinating modality-specific simulation processes through consistent behavioral states and physical controls.

\subsubsection{UPR State Initialization and Evolution}

During generation and interactive refinement, the UPR is maintained as a constrained behavioral state representation updated by reasoning agents according to multimodal inputs and simulation objectives. Given an initial UPR state $\mathbf{p}(t)$, multimodal observations and semantic interpretations $\mathcal{Z}(t)$, and global objectives $\mathcal{G}$, the Orchestrator Agent computes state updates:
\begin{equation}
\mathcal{F}: (\mathcal{Z}(t),\mathbf{p}(t),\mathcal{G})
\rightarrow
\Delta\mathbf{p},
\end{equation}
where $\mathcal{G}$ specifies constraints including semantic consistency, multi-agent coordination, and interactive requirements.

The update consists of two components. Observation-conditioned updates,
\begin{equation}
\Delta\mathbf{p}_{H}
=
\mathcal{H}(\mathcal{Z}(t),\mathbf{p}(t)),
\end{equation}
capture changes introduced by new multimodal inputs and semantic interpretation. Goal-conditioned updates,
\begin{equation}
\Delta\mathbf{p}_{\Omega}
=
\Omega(\mathbf{p}(t),\mathcal{G}),
\end{equation}
maintain consistency among simulated agents, behavioral states, and environmental conditions.

The complete update is:
\begin{equation}
\Delta\mathbf{p}
=
\Delta\mathbf{p}_{H}
+
\Delta\mathbf{p}_{\Omega},
\end{equation}
followed by projection into a valid state space:
\begin{equation}
\mathbf{p}(t+\Delta t)
=
\Pi_{\mathcal{P}_{\mathrm{valid}}}
(\mathbf{p}(t)+\Delta\mathbf{p}),
\end{equation}
where $\Pi_{\mathcal{P}_{\mathrm{valid}}}$ enforces semantic, physical, and cross-modal constraints before the updated UPR state is interpreted by physically-based simulation agents.

In the current implementation, $\mathcal{H}$ and $\Omega$ are realized through structured prompting strategies and rule-based constraints rather than learned state dynamics.

\subsection{Input Layer}

The Input Layer receives multimodal instructions, including text, images, and optional audio descriptions. Inputs may specify object types, behavioral states, affective expressions, environmental conditions, and interaction scenarios. The system first extracts structured semantic attributes, which are then transformed into behavioral representations and physical control parameters through the reasoning and orchestration layers.

\subsection{Reasoning and Orchestration Layer}

The Reasoning and Orchestration Layer performs multimodal interpretation, behavioral reasoning, agent coordination, and semantic-to-physical transformation through two classes of specialized LLM-based agents: the Orchestrator Agent and Reasoning Agents.

\subsubsection{Orchestrator Agent}

The Orchestrator Agent coordinates interactions among reasoning agents and physically-based simulation agents, maintaining consistency across semantic states, behavioral representations, and execution parameters by updating the shared UPR throughout initial generation and interactive refinement.

\subsubsection{Reasoning Agents}

The LLM-based reasoning agents operate on multimodal inputs and maintain the shared UPR representation that connects high-level semantic intent with physically executable simulation parameters.

\paragraph{Interpretation Agent.}

The Interpretation Agent analyzes multimodal inputs and extracts semantic attributes, including type identity, environmental context, behavioral intent, affective state, and interaction information. These attributes form the semantic representation used for downstream transformation.

\paragraph{Transformation Agent.}

The Transformation Agent performs semantic-to-physical mapping by converting semantic and behavioral representations into interpretable kinematic and acoustic parameters stored in the UPR. These parameters are subsequently executed by physically-based simulation agents.

\subsubsection{Semantic-to-Physical Mapping}
\label{sec:semantic_physical_mapping}

The proposed architecture performs explicit semantic-to-physical transformation through the UPR rather than directly generating motion and sound from multimodal inputs. Semantic attributes are represented as:
\begin{equation}
\mathcal{P}_{sem}
=
\{
s_{\text{type}},
s_{\text{behavior}},
s_{\text{emotion}},
s_{\text{interaction}}
\},
\end{equation}

and environmental attributes are represented as:
\begin{equation}
\mathcal{P}_{env}
=
\{
e_{\text{scene}},
e_{\text{time}},
e_{\text{weather}}
\}.
\end{equation}

These attributes are transformed into a shared behavioral state within the UPR:
\begin{equation}
\mathcal{P}_{beh}
=
F(\mathcal{P}_{sem},\mathcal{P}_{env}),
\end{equation}
where $\mathcal{P}_{beh}$ encodes behavior-level properties, including activity intensity, movement style, vocalization tendency, and interaction patterns.

The shared behavioral state is then interpreted by modality-specific simulation agents to derive execution parameters:
\begin{equation}
\mathcal{P}_{kin}
=
G_{kin}
(
\mathcal{P}_{beh},
\mathcal{P}_{sem},
\mathcal{P}_{env}
),
\end{equation}
\begin{equation}
\mathcal{P}_{acous}
=
G_{acous}
(
\mathcal{P}_{beh},
\mathcal{P}_{sem},
\mathcal{P}_{env}
).
\end{equation}

The resulting kinematic and acoustic parameters provide a compact, interpretable control representation for the physically-based simulation agents, governing flight speed, trajectory variation, vocalization rate, frequency characteristics, and spatial sound properties. Through the joint representation of motion and acoustics, the UPR enables the simulation agents to generate coordinated 3D behaviors and spatialized soundscapes while preserving consistency with the original multimodal intent. The low-dimensional parameterization provides direct control over both kinematic and acoustic behaviors, supporting synchronized and expressive simulation.

\subsection{Execution Layer}

The Execution Layer translates the shared UPR state into coordinated multimodal outputs through specialized physics-based simulation agents. Each agent derives modality-specific control parameters from the UPR and executes them using interpretable computational models with explicit control over kinematic, acoustic, and environmental variables. These agents directly execute the UPR without semantic reasoning or decision-making, producing physically grounded, reproducible, and cross-modally consistent simulations.

\subsubsection{Kinematic Agent}
\label{sec:Kinematic_agent}

The Kinematic Agent generates physically grounded 3D trajectories using kinematic parameters derived from the UPR state. The motion model uses stochastic continuous-time dynamics  to produce smooth and controllable flight behaviors with plausible variations \cite{10.1145/37401.37406}. Species characteristics, behavioral states, and environmental conditions are encoded through the kinematic parameters, providing diverse and interpretable motion generation.

\paragraph{Position} Each flying or moving object is modeled as a stochastic self-propelled particle in a
three-dimensional environment. The state of object $i$ at time $t$ is
represented by its position
\begin{equation}
\mathbf{x}_i(t)
=
\begin{bmatrix}
x_i(t)\\
y_i(t)\\
z_i(t)
\end{bmatrix},
\end{equation}
where $\mathbf{x}_i(t)$ denotes the object's position in world coordinates.

\paragraph{Velocity}
The translational velocity of object $i$ is represented by its horizontal speed, horizontal heading, and vertical speed:
% \begin{equation}
% \dot{\mathbf{x}}_i(t) =
% v_i(t)
% \begin{bmatrix}
% \cos\psi_i(t)\cos\varphi_i(t) \\
% \cos\psi_i(t)\sin\varphi_i(t) \\
% \sin\psi_i(t)
% \end{bmatrix}.
% \end{equation}
\begin{equation}
\dot{\mathbf{x}}_i(t)
=
\begin{bmatrix}
v_i(t)\cos\varphi_i(t)\\
v_i(t)\sin\varphi_i(t)\\
v_{z,i}(t)
\end{bmatrix},
\end{equation}
where $v_i(t)$ denotes the horizontal speed, $\varphi_i(t)$ specifies the horizontal heading (azimuth), and $v_{z,i}(t)$ denotes the vertical speed. Thus, the translation motion state is characterized by $\varphi_i(t)$, $v_i(t)$, and $v_{z,i}(t)$. This formulation determines the continuous position evolution while allowing stochastic motion in three-dimensional space \cite{10.1145/37401.37406}. The motion dynamics and orientation dynamics evolve according to stochastic Ornstein--Uhlenbeck process~\cite{UhlenbeckOrnstein30}.

The motion state is controlled by the UPR kinematic representation $\mathcal{P}_{kin}$ with parameter definitions and behavioral effects given in Table~\ref{tab:upr_control_params}.

\subsubsection{Acoustic Agent}
\label{sec:acoustic_agent}
The acoustic agent generates procedural bird-like vocalizations and multi-object spatialized soundscapes \cite{11402190} through signal-based synthesis and spatial acoustic modeling \cite{oppenheim2010discrete}. 
The agent directly controls acoustic, temporal, and spatial properties through interpretable variables, with $\mathcal{P}_{acous}$ derived from semantic, behavioral, species, and environmental attributes and coordinated with the UPR to ensure consistency between vocal behavior and simulated motion.

\paragraph{3D Spatialization} Each object $i$ emits a sequence of chirps, where $c_{i,k}(t)$ denotes the waveform of the $k$-th chirp emitted by object $i$ at time $t$. The chirp is parameterized by species-specific acoustic features. Each vocal event is spatialized along the corresponding object trajectory $\mathbf{x}_i(t)$:
\begin{equation}
\mathbf{s}_{i,k}(t)
=
A\!\left(\mathbf{x}_i(t)\right)c_{i,k}(t),
\end{equation}
where $A\!\left(\mathbf{x}_i(t)\right)$ denotes the spatial acoustic operator for distance attenuation, directional gain, and stereo localization. Independent trajectories produce independently positioned vocal sources, forming a synchronized multi-agent 3D sound field.

\subsubsection{Rendering Agent}

The Rendering Agent generates environment-aware visual scenes from $\mathcal{P}_{env}$ and simulated object trajectories. Environmental attributes, including habitat, lighting, and weather conditions, are synchronized with motion and acoustic generation to maintain consistency across visual, auditory, and behavioral modalities.

\subsubsection{Kinematic and Acoustic Control Parameters}
Table~\ref{tab:upr_control_params} summarizes the kinematic and acoustic parameters forming a compact, interpretable representation for coordinated 3D motion and spatialized soundscapes.

\begin{table}[t]
\footnotesize
\centering
\caption{Kinematic and Acoustic Control Parameters}
\begin{tabular}{lll}
\toprule
\textbf{Parameter} & \textbf{Description} & \textbf{Behavioral Effect} \\
\midrule
\multicolumn{3}{l}{\textit{Kinematic}} \\
$\vec{x}_0$ & Equilibrium position & Nominal CoM location \\
$\Delta \vec{x}$ & Oscillation amplitudes & Path curvature \\
$\lambda$ & Relaxation rates & Agility \\
$\sigma$ & Noise amplitudes & Stochastic variability \\
$v_0$ & Preferred horizontal speed & Target forward velocity \\
$v_{z,0}$ & Preferred vertical speed & Target climb/descent rate \\
$\psi_0$ & Equilibrium pitch & Nominal body orientation \\
\midrule
\multicolumn{3}{l}{\textit{Acoustic}} \\
$f_0,f_1$ & Chirp start/end frequencies & Frequency sweep range \\
$d$ & Chirp duration & Length of vocalization \\
$\alpha$ & Trill strength & Pitch oscillation \\
$f_\text{trill}$ & Trill rate & Vibrato speed \\
$A(\mathbf{p})$ & Spatial gain & Stereo localization \\
\bottomrule
\end{tabular}
\label{tab:upr_control_params}
\end{table}

%-----------------------------
%%%%%%%%%%%%%%%%%%%%%%%%%%%%%%%%%%%%%%%%%%%%%%%%%%%%%
%----------------------------------
% Table: Category Definitions
%----------------------------------
\begin{table*}[t]
\footnotesize
\centering
\caption{Category Definitions}
\begin{tabular}{p{2cm} p{6cm} p{8.5cm}}
\hline
\textbf{Dimension} & \textbf{Values} & \textbf{Description} \\
\hline
Bird Species & owl, nightingale, lark, warbler, sparrow, finch, raven & Distinct vocal patterns for species-specific acoustic assessment \\
Environments & night, moon, moonlight, daytime, sun, forest, storm, rain, wind, sand, home & Guides scene composition and 3D auditory spatialization \\
Emotional States & happy, fear, sad & Modulates motion (speed, trajectory) and sound (pitch, rhythm, intensity) \\
\hline
\end{tabular}
\label{tab:input_dimensions}
\end{table*}
%%%%%%%%%%%%%%%%%%%%%%%%%%%%%%%%%%%%%%%%%%%%%%%%%%%%%

%%%%%%%%%%%%%%%%%%%%%%%%%%%%%%%%%%%%%%%%%%%%%%%%%%%%%
%----------------------------------
% Table: Category Definitions
%----------------------------------
% \begin{table}[t]
% \footnotesize
% \centering
% \caption{Category Definitions}
% \begin{tabular}{p{1.5cm} p{2.5cm} p{2.5cm}}
% \hline
% \textbf{Dimension} & \textbf{Values} & \textbf{Description} \\
% \hline
% Bird Species & owl, nightingale, lark, warbler, sparrow, finch, raven & Distinct vocal patterns for species-specific acoustic assessment \\
% Environments & night, moon, moonlight, daytime, sun, forest, storm, rain, wind, sand, home & Guides scene composition and 3D auditory spatialization \\
% Emotional States & happy, fear, sad & Modulates motion (speed, trajectory) and sound (pitch, rhythm, intensity) \\
% \hline
% \end{tabular}
% \label{tab:input_dimensions}
% \end{table}
%%%%%%%%%%%%%%%%%%%%%%%%%%%%%%%%%%%%%%%%%%%%%%%%%%%%%
%-----------------------------
%-----------------------------
%%%%%%%%%%%%%%%%%%%%%%%%%%%%%%%%%%%%%%%%%%%%%%%%%%%%%
%-------------------------------
% Table: Score Definitions
%-------------------------------
\begin{table*}[h]
\footnotesize
\centering
\caption{Score definitions for cross-modal coherence, affective consistency, and generation consistency (0--3 scale).}
\begin{tabular}{c p{4.8cm} p{5.2cm} p{5.5cm}}
% \begin{tabular}{c p{5.5cm} p{4.5cm} p{4cm}}
\hline
\textbf{Score} & \textbf{Cross-Modal Coherence} & \textbf{Affective Consistency} & \textbf{Generation Consistency} \\
\hline
3 & Strong motion, sound, and environment alignment; physically plausible & Behavior and sound match the specified affective descriptor & Stable outputs across runs with natural variation \\
2 & Minor mismatch but overall coherent & Partial affective correspondence & Noticeable variation with preserved semantics \\
1 & Major mismatch or implausible behavior & Weak or unclear affective correspondence & Large variation affecting consistency \\
0 & Failed multimodal generation & No affective correspondence & Inconsistent outputs across runs \\
\hline
\end{tabular}
\label{tab:score_def}
\end{table*}
%%%%%%%%%%%%%%%%%%%%%%%%%%%%%%%%%%%%%%%%%%%%%%%%%%%%%
%-----------------------------

\subsection{Output Layer and Interactive Loop}

The Output Layer integrates the outputs of the physically-based simulation agents into a synchronized multimodal scene containing 3D motion, spatialized birdsong, and environment-aware rendering.

The framework supports iterative interaction through a closed-loop process. Users can modify species, behaviors, affective states, or environmental conditions through additional multimodal instructions. The LLM-based reasoning agents update semantic and behavioral representations, the Orchestrator Agent propagates these changes through the UPR, and then the physically-based simulation agents regenerate the resulting multimodal outputs.

This architecture maintains an interpretable connection between LLM-based reasoning agents and physically-based simulation agents through the UPR. By decoupling semantic reasoning from physical execution, it provides explicit control over semantic, behavioral, kinematic, acoustic, and environmental states, forming a general framework for interactive multimodal embodied behavior generation.

%============================================
%%-----------------------------------------------------

\section{Experiments and Evaluations}

We develop \emph{BirdsongChat} as a prototype system to implement and evaluate the proposed hybrid multi-agent framework, which integrates reasoning agents with physics-based simulation agents to generate coordinated multi-bird motion, spatialized birdsong soundscapes, and environment-aware scenes from multimodal inputs. The evaluation examines semantic alignment, cross-modal coherence, interactive refinement, and simulation consistency across text- and image-guided scenarios.

\subsection{Experimental Setup}
\label{subsec:exp_setup}

The BirdsongChat prototype is deployed on Amazon Web Services (AWS). For these experiments, the prototype uses the Amazon Nova 2 multimodal AI foundation model to support reasoning agents for multimodal interpretation, orchestration, and semantic-to-physical transformation; however, the model can be interchanged with any other model.

The evaluation covers combinations of bird species, behavioral descriptions, affective states, and environmental conditions represented in the UPR. Table~\ref{tab:input_dimensions} summarizes the evaluated input categories.

%-----------------------------
%%%%%%%%%%%%%%%%%%%%%%%%%%%%%%%%%%%%%%%%%%%%%%%%%%%%%
%---------------------------------------------------
% Table: Text prompt alignment effectiveness scores
%---------------------------------------------------
\begin{table*}[h]
\footnotesize
\centering
\caption{Text prompt alignment effectiveness with per-sample scores (0--3 scale).}
\begin{tabular}{c p{5cm} p{2.5cm} p{1cm} p{2.7cm} c c c c}
\hline
\textbf{ID} & \textbf{Text Prompt} & \textbf{Environment} & \textbf{Affect} & \textbf{Species} & \textbf{Correct} & \textbf{Coh.} & \textbf{Aff.} & \textbf{Gen.} \\
\hline
1 & The moonlight shines brightly on the damp forest floor. & Night, Moon, Forest & Fear/Sad & Nightingale, Owl & Yes & 3 & 3 & 3 \\
2 & The backyard forest sings harmonically with the newly bloomed flowers. & Daytime, Sun, Forest & Happy & Sparrow, Finch, Raven & Yes & 3 & 3 & 3 \\
3 & The birds sing happily with another under the gleaming sun. & Sun, Wind, Forest & Happy & Sparrow, Finch, Raven & Yes & 3 & 3 & 3 \\
4 & In a gloomy dark night, birds chirp silently in the forbidden forest of fear. & Night, Forest & Fear & Nightingale, Owl & Yes & 3 & 3 & 3 \\
5 & The birds shudder as a haunting call echoes from the distance. & Night, Moon & Fear & Owl, Nightingale & Yes & 3 & 3 & 3 \\
6 & Everyone in the forest chirps excitedly as the new dawn comes. & Sun, Wind, Forest & Happy & Sparrow, Raven, Finch & Yes & 3 & 3 & 3 \\
7 & The forest is ravaged by a storm, the birds displaced from their homes. & Night, Forest, Wind & Sad & Finch & Yes & 3 & 3 & 3 \\
8 & The birds freak out and scatter away. & Storm, Wind, (Forest)& Fear & Raven, Sparrow & Yes & 3 & 3 & 2 \\
9 & The birds danced on the dunes, feeling elated that the sunlight has finally come. & Sun, Wind, (Sand)  & Happy & Lark & Yes & 3 & 3 & 2 \\
10 & The birds all mourn in silence at night, the passings still fresh in their minds. & Night, Moon, Forest & Sad & Nightingale, Owl  & Yes & 3 & 3 & 3 \\
11 & The birds cower from the shadows looming over the forest. & Night, Forest & Fear/Sad & Owl, Nightingale & Yes & 3 & 3 & 3 \\
12 & Under the sunlight, the birds fly around animatedly. & Sun, Wind, (Forest) & Happy & Sparrow, Raven, Finch & Yes & 3 & 3 & 3 \\
\hline
\end{tabular}
\label{tab:text_scored}
\end{table*}
%%%%%%%%%%%%%%%%%%%%%%%%%%%%%%%%%%%%%%%%%%%%%%%%%%%%%

%%%%%%%%%%%%%%%%%%%%%%%%%%%%%%%%%%%%%%%%%%%%%%%%%%%%%
%---------------------------------------------------
% Table: Image prompt alignment effect scores
%---------------------------------------------------
\begin{table*}[h]
\footnotesize
\centering
\caption{Image prompt evaluation results with per-sample scores (0--3 scale).}
\begin{tabular}{c p{4cm} p{2.3cm} p{1.2cm} p{3cm} c c c}
\hline
\textbf{ID} & \textbf{Image Context} & \textbf{Environment} & \textbf{Affect} & \textbf{Species} & 
\textbf{Coh.} & \textbf{Aff.} & \textbf{Gen.}\\
\hline
1 & Sunset & Sun, Forest, Storm & Happy & Raven, Sparrow, Finch & 2 & 3 & 2\\
2 & Beach with sand and sea & Sun, Sand, Daytime & Happy & Lark & 3 & 3 & 3\\
3 & Beach with sand, sea, and pine tree & Sun, Sand & Happy & Lark & 3 & 3 & 3\\
4 & Nighttime moon and sea & Night, Moon, Sea & Fear & Nightingale, Owl & 3 & 3 & 3\\
5 & Thunderstorm & Storm, Wind & Fear & Raven, Sparrow & 3 & 3 & 3\\
6 & Foggy golf course & Fog, Forest & Happy & Nightingale, Owl & 1 & 3 & 2\\
\hline
\end{tabular}
\label{tab:image_scored}
\end{table*}

%%%%%%%%%%%%%%%%%%%%%%%%%%%%%%%%%%%%%%%%%%%%%%%%%%%%%
%-------------------------------
% Figure: Image Prompt Samples
%-------------------------------
\begin{figure*}[h]
    \centering

% define a fixed width/height for all subfigures using minipage
    \newlength{\subfigsize}
    \setlength{\subfigsize}{0.15\textwidth}  % adjust as needed
 \begin{subfigure}[t]{\subfigsize}
  \centering
  \includegraphics[width=\textwidth]{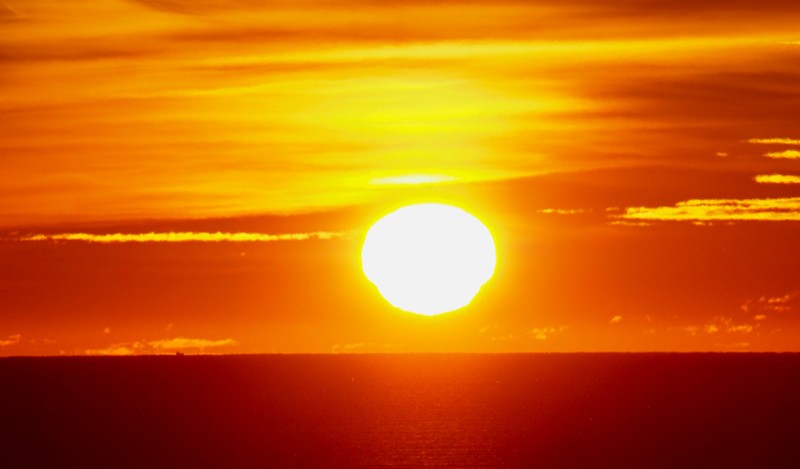}
  \caption{\textnormal{Image 1\\}}
  \end{subfigure}
  \hfill
   \begin{subfigure}[t]{\subfigsize}
  \centering
  \includegraphics[width=\textwidth]{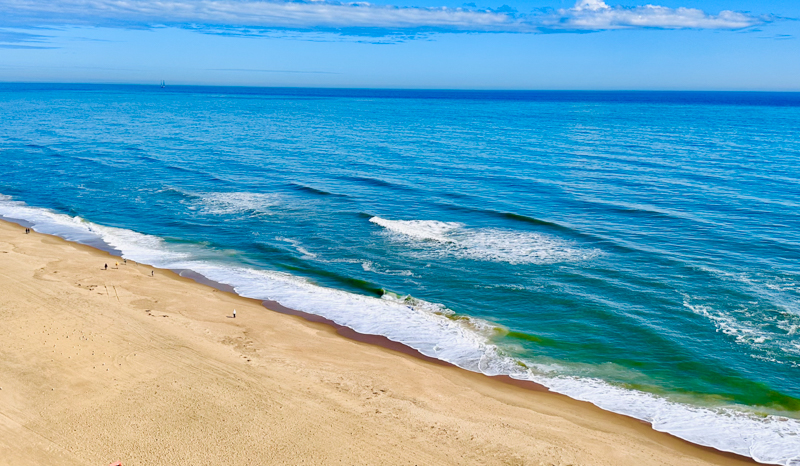}
  \caption{\textnormal{Image 2}}
 \end{subfigure}
 \hfill
 \begin{subfigure}[t]{\subfigsize}
  \centering
  \includegraphics[width=\textwidth]{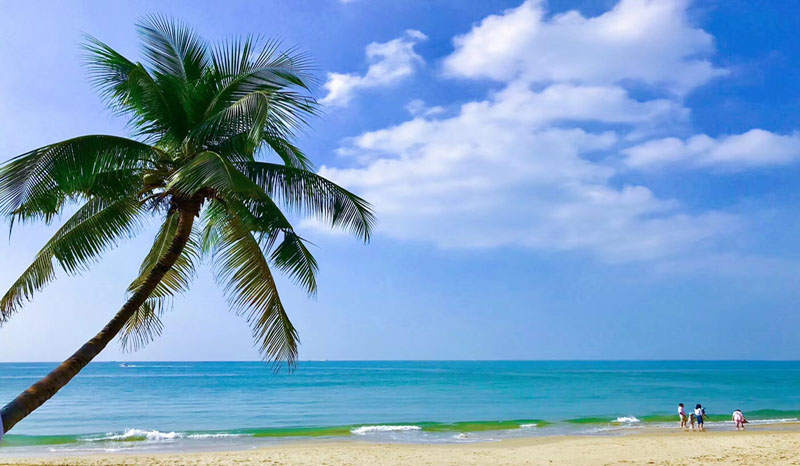}
  \caption{\textnormal{Image 3}}
 \end{subfigure}
 \hfill
  \begin{subfigure}[t]{\subfigsize}
  \centering
  \includegraphics[width=\textwidth]{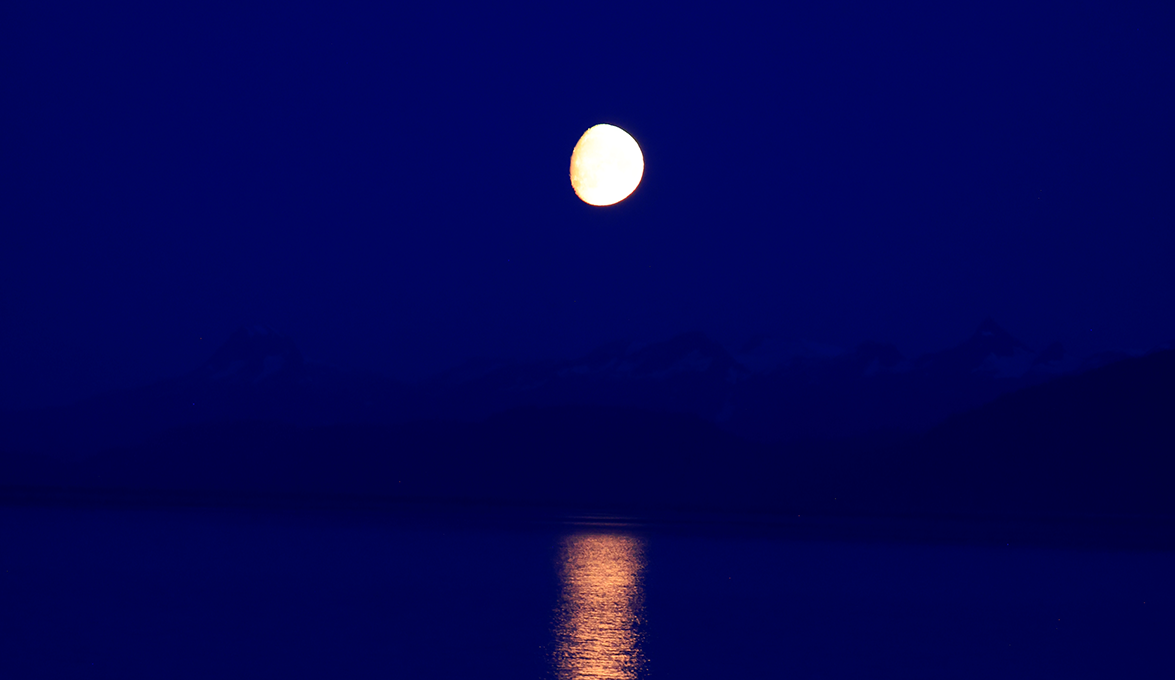}
  \caption{\textnormal{Image 4}}
 \end{subfigure}
 \hfill
  \begin{subfigure}[t]{\subfigsize}
  \centering
  \includegraphics[width=\textwidth]{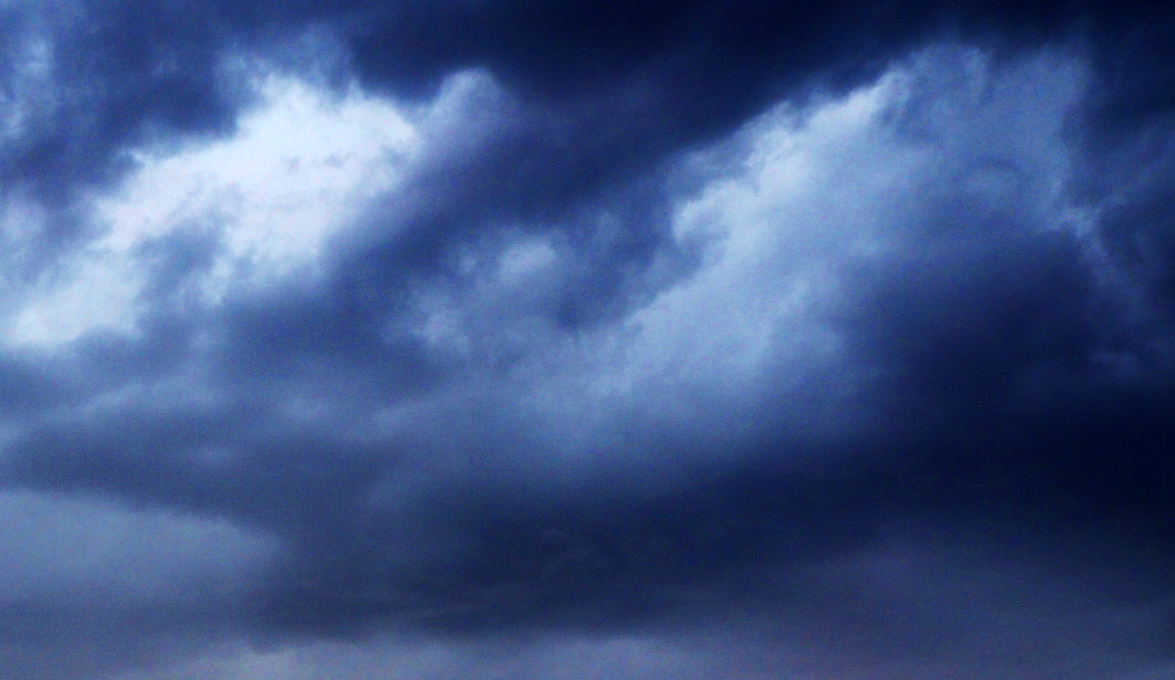}
  \caption{\textnormal{Image 5}}
 \end{subfigure}
 \hfill
  \begin{subfigure}[t]{\subfigsize}
  \centering
  \includegraphics[width=\textwidth]{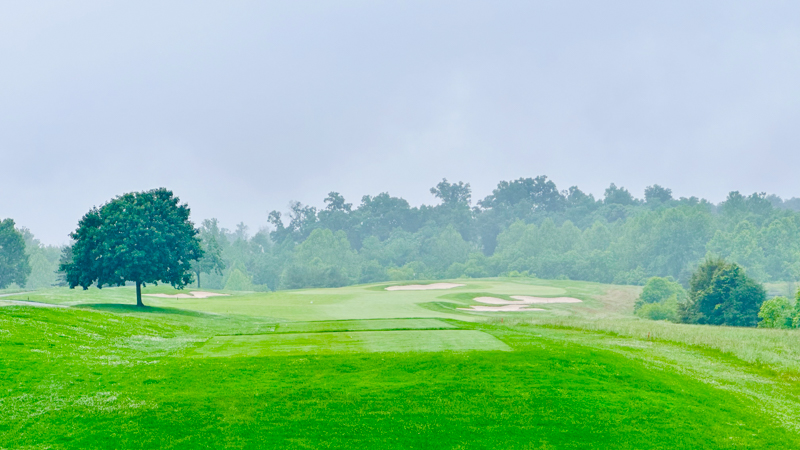}
  \caption{\textnormal{Image 6}}
 \end{subfigure}
 
 \caption{Experimental image prompt examples.}
 \label{fig:image_inputs}
\end{figure*}
%%%%%%%%%%%%%%%%%%%%%%%%%%%%%%%%%%%%%%%%%%%%%%%%%%%%%

%--------------------------------

%%%%%%%%%%%%%%%%%%%%%%%%%%%%%%%%%%%%%%%%%%%%%%%%%%%%%%%%%%%%%
%%% Figures: BirsongChat Demo Groups
%%%%%%%%%%%%%%%%%%%%%%%%%%%%%%%%%%%%%%%%%%%%%%%%%%%%%%%%%%%%%
\begin{figure*}[h]
    \centering
    
 \begin{subfigure}[t]{0.48\textwidth}
  \centering
 \begin{mdframed}[linecolor=gray, linewidth=0.5pt, roundcorner=10pt]
 \includegraphics[width=\textwidth]{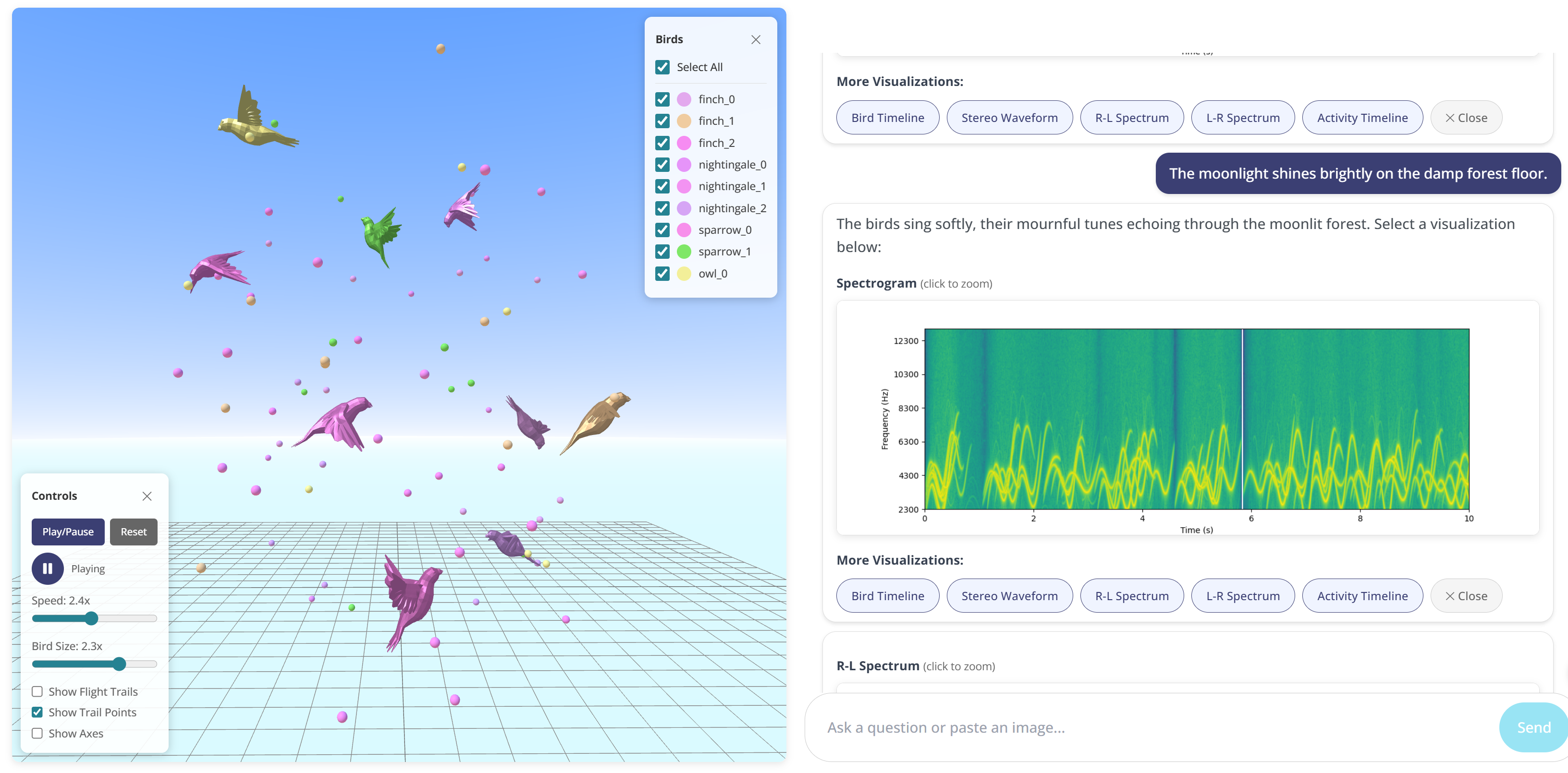}
 \end{mdframed}
  \caption{\textnormal{
BirdsongChat generates birdsong with 3D flight trajectories and spectrograms, visualizing bird species, routes, and trail points. A melancholic nighttime prompt produces slower, lower-frequency sounds featuring nocturnal species such as owls and nightingales. Only trail points are shown for a clearer view of bird positions.\\}}
  \end{subfigure}
  \hfill
  \begin{subfigure}[t]{0.48\textwidth}
  \centering
  \begin{mdframed}[linecolor=gray, linewidth=0.5pt, roundcorner=10pt]
  \includegraphics[width=\textwidth]{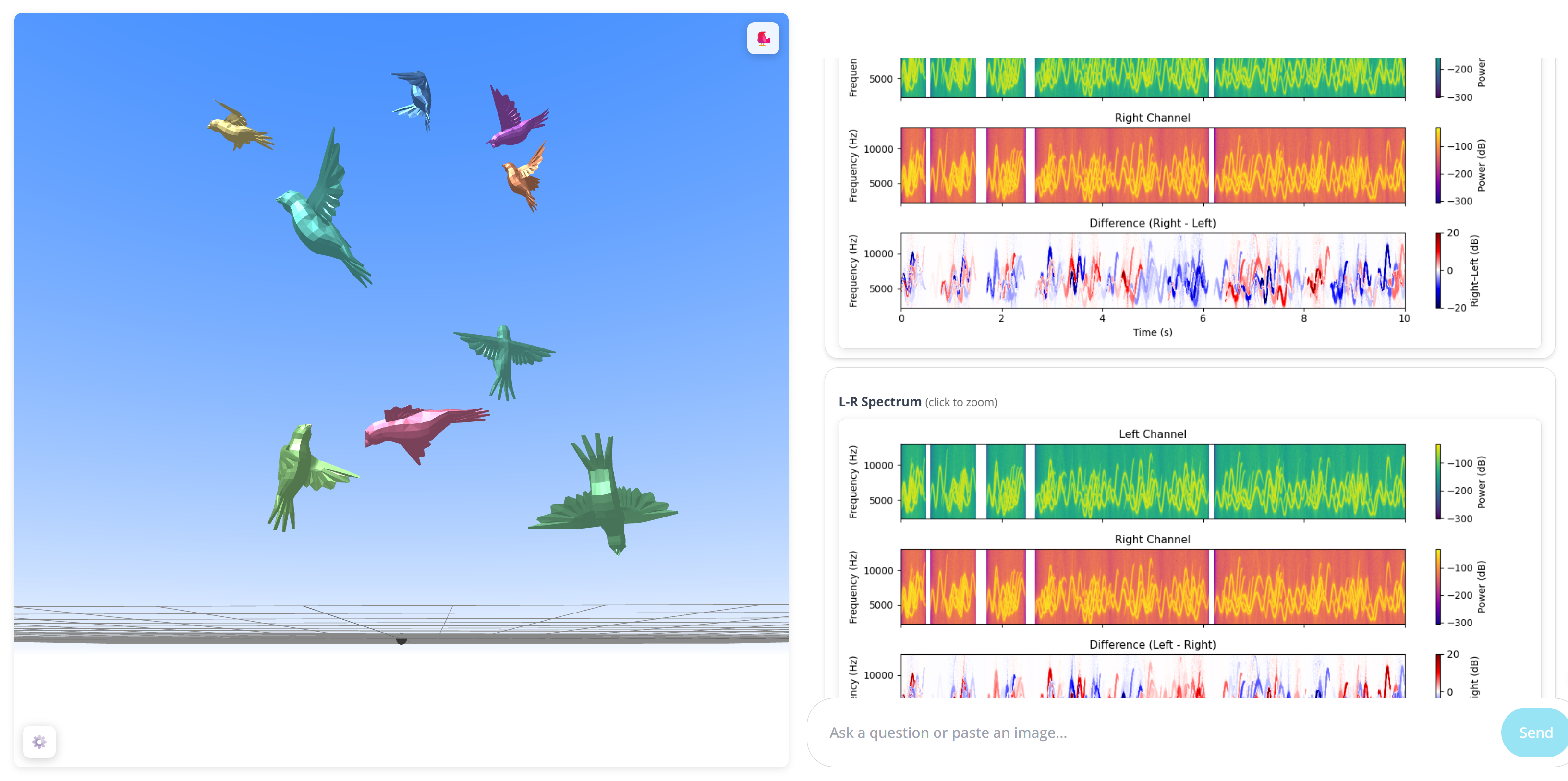}
 \end{mdframed}
  \caption{\textnormal{The display shows the L–R spectrum, with the lower graph highlighting frequency differences and left–right channel bias. Collapsed GUI sliders slightly enlarge the bird objects. Flight trails, trail points, and axes are hidden.}}
 \end{subfigure}

 \begin{subfigure}[t]{0.48\textwidth}
 \centering
 \begin{mdframed}[linecolor=gray, linewidth=0.5pt, roundcorner=10pt]
 \includegraphics[width=\textwidth]{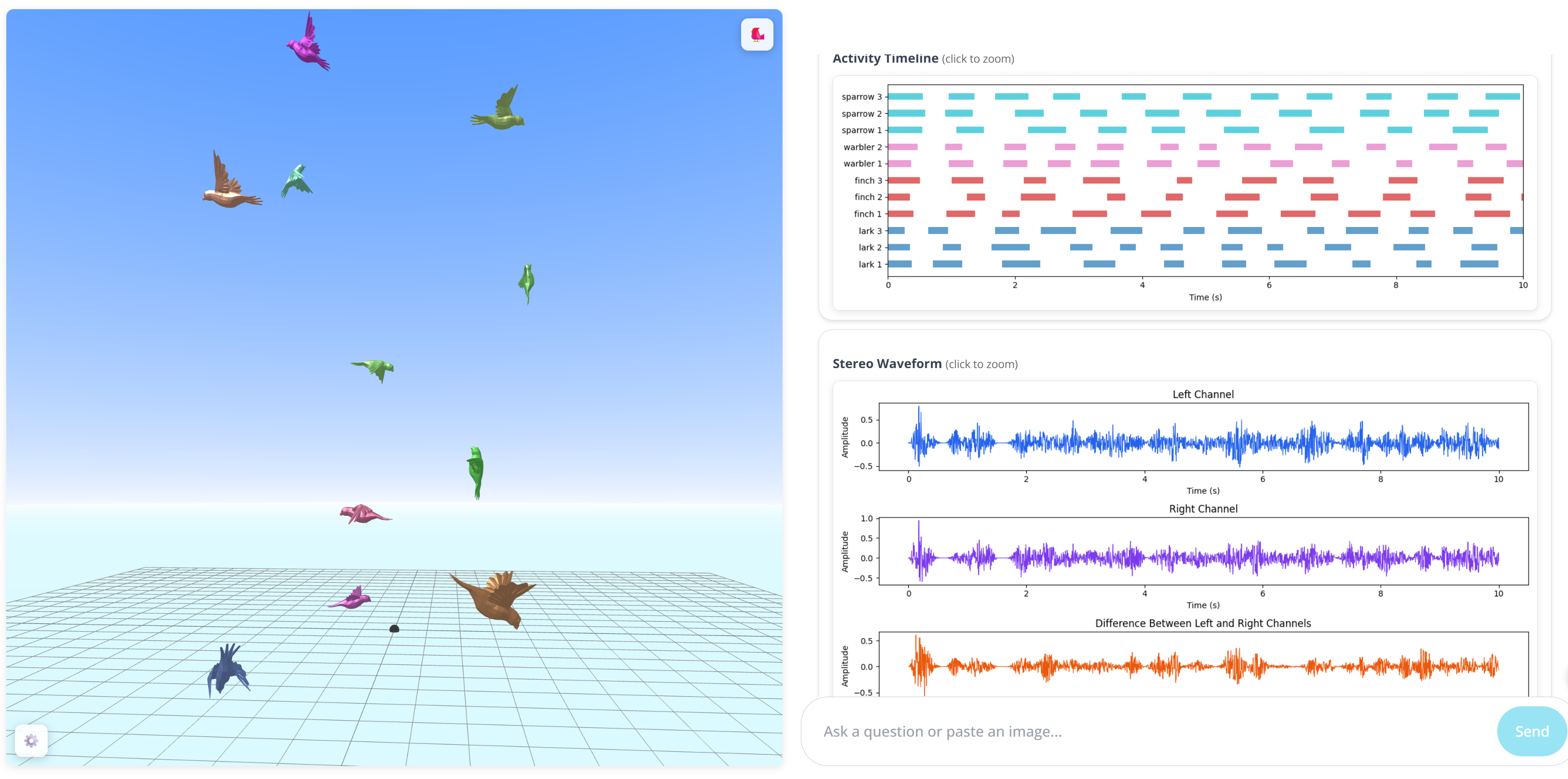}
 \end{mdframed}
  \caption{\textnormal{This display shows the birdsong activity timeline and stereo waveforms. The lower graph highlights channel differences, while the timeline marks bird calls by species and color. Call patterns vary, with warblers producing shorter, more consistent calls than finches. Only the birds are shown to simulate a natural flight scene.\\}}
 \end{subfigure}
  \hfill
  \begin{subfigure}[t]{0.48\textwidth}
  \centering
 \begin{mdframed}[linecolor=gray, linewidth=0.5pt, roundcorner=10pt]
 \includegraphics[width=\textwidth]{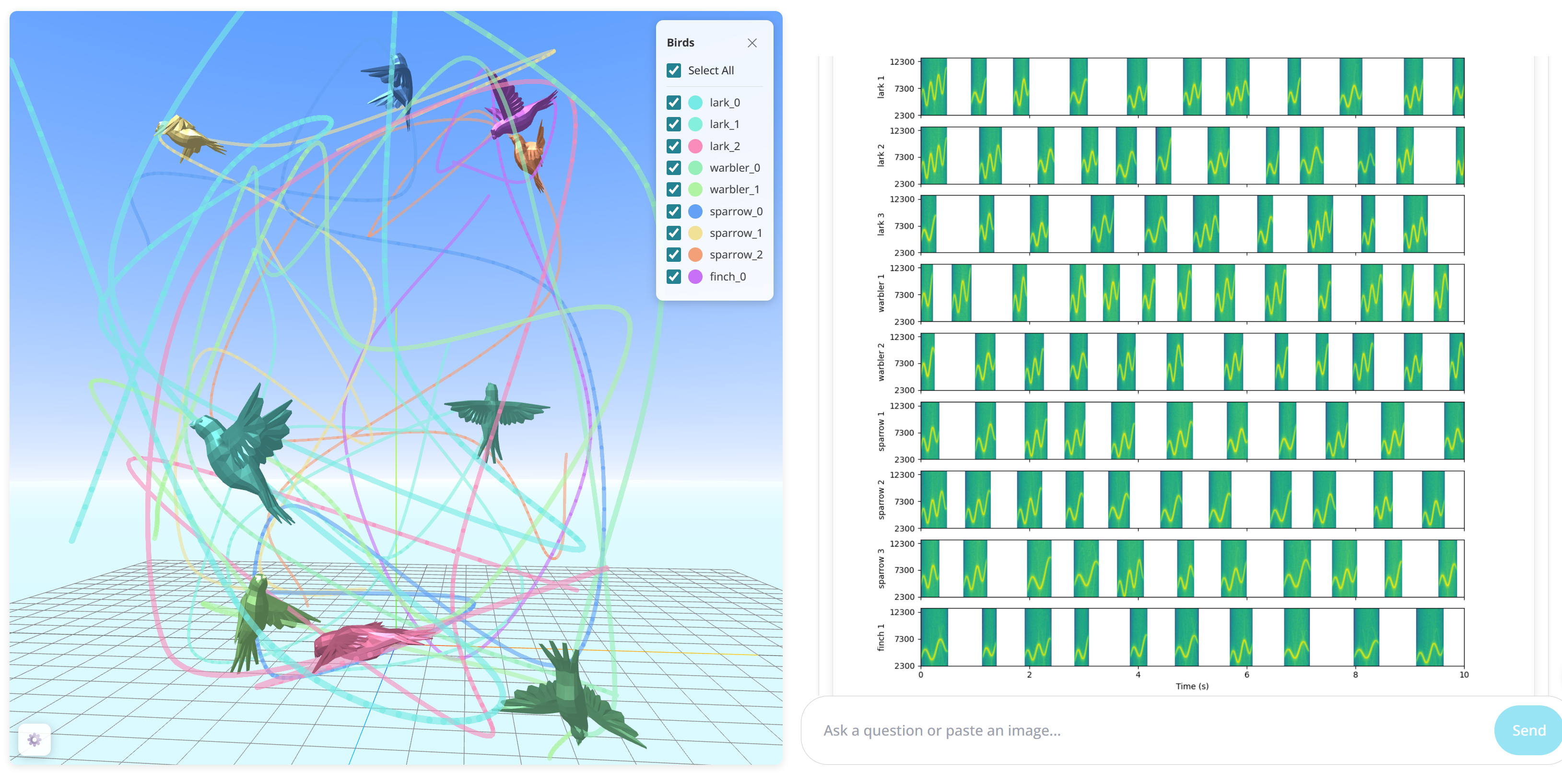}
 \end{mdframed}
  \caption{\textnormal{This display shows bird frequencies, pitches, and call timing. Higher pitches suggest positive sentiment, while lower pitches suggest negative sentiment. Here, fast, high-frequency calls reflect a positive mood. Enlarged birds and flight trails highlight each bird’s calculations.}}
 \end{subfigure}
 
 \caption{BirdsongChat interface and examples. Users can directly copy prompts to try the AWS web-based prototype system.}
 \label{fig:demogroup_4pics}
\end{figure*}

\begin{table*}[t]
\centering
\footnotesize
\setlength{\tabcolsep}{12pt}
\caption{Quantitative evaluation results (0--3 scale). Each prompt is executed five times to evaluate semantic alignment and generation consistency.}
\label{tab:eval_quan_all}

\begin{tabular}{lcccc}
\hline
\textbf{Metric} & \textbf{Average} & \textbf{Min} & \textbf{Max} & \textbf{Score (\%)} \\
\hline

\multicolumn{5}{l}{\textit{\textbf{Text Prompts} (12 prompts, 60 runs)}}\\
Cross-modal Coherence & 3.00 & 3 & 3 & 100\%\\
Affective Consistency & 3.00 & 3 & 3 & 100\%\\
Generation Consistency & 2.83 & 2 & 3 & 94.4\%\\
\textbf{Overall} & \textbf{2.94} & \textbf{2} & \textbf{3} & \textbf{98.1\%}\\

\hline

\multicolumn{5}{l}{\textit{\textbf{Image Prompts} (6 prompts, 30 runs)}}\\
Cross-modal Coherence & 2.50 & 1 & 3 & 83.3\%\\
Affective Consistency & 3.00 & 3 & 3 & 100\%\\
Generation Consistency & 2.67 & 2 & 3 & 88.9\%\\
\textbf{Overall} & \textbf{2.72} & \textbf{1} & \textbf{3} & \textbf{90.7\%}\\

\hline

\multicolumn{5}{l}{\textit{\textbf{Overall Evaluation} (18 prompts, 90 runs)}}\\
Cross-modal Coherence & 2.83 & 1 & 3 & 94.4\%\\
Affective Consistency & 3.00 & 3 & 3 & 100\%\\
Generation Consistency & 2.78 & 2 & 3 & 92.6\%\\
\textbf{Overall} & \textbf{2.87} & \textbf{1} & \textbf{3} & \textbf{95.7\%}\\

\hline
\end{tabular}
\label{tab:quant_eval_results_detail}
\end{table*}
%-----------------------------

\subsection{Qualitative Evaluation}

BirdsongChat is evaluated across representative scenarios to assess whether the interaction between LLM-based reasoning agents and physics-based simulation agents maintains semantic consistency, behavioral plausibility, and cross-modal alignment among generated motion, sound, and environmental context (Figures~\ref{fig:demogroup_4pics}). The evaluation focuses on affective behavior mapping, environmental grounding, multimodal consistency, stochastic variation, multi-bird coordination, and interactive refinement.

\emph{Affective Behavior Mapping.}
The LLM-based reasoning agents interpret affective descriptors such as \emph{happy}, \emph{sad}, and \emph{fear} and transform them into behavioral representations within the UPR. Physically-based simulation agents subsequently convert these representations into coordinated motion and acoustic variations. High-activity descriptors result in increased flight speed, trajectory variation, and denser vocalization patterns, whereas low-activity descriptors produce slower motion and sparser acoustic events. These results demonstrate the effectiveness of the UPR as an intermediate semantic-to-physical interface.

\emph{Environment-Conditioned Behaviors.}
Multimodal inputs, including text and images, provide environmental context that influences semantic interpretation, behavioral representation, and scene generation. Different habitats and temporal conditions lead to corresponding variations in simulated species, motion patterns, acoustic characteristics, and environmental rendering. These results demonstrate that contextual information is propagated from reasoning agents to physically-based simulation agents through the UPR.

\emph{Cross-Modal Alignment.}
Motion and acoustic generation are coordinated through the shared UPR representation. Changes in simulated bird trajectories influence spatial sound properties, including localization, attenuation, and temporal relationships among multiple vocal sources. The resulting audiovisual outputs maintain consistency among simulated motion, acoustic events, and environmental context.

\emph{Stochastic Variation and Simulation Stability.}
The physically-based simulation agents incorporate bounded stochastic processes to introduce natural variation while preserving controllability. Stochastic motion dynamics and acoustic modulation generate diverse trajectories and vocalization patterns without significant deviation from the intended semantic conditions. This enables repeatable simulation while maintaining behavioral diversity.

\emph{Multi-Bird Control and Interactive Refinement.}
The framework supports simultaneous simulation of multiple birds and species within shared environments. Through the UPR interface, users can modify species, behaviors, affective descriptors, and environmental conditions using conversational commands or image prompts. These updates are propagated from reasoning agents to simulation agents while preserving consistency across motion, sound, and rendering.

\subsection{Quantitative Evaluation}

\subsubsection{Quantitative Evaluation Protocol}

Representative scenarios cover combinations of bird species, environmental contexts, and affective descriptors defined in Section~\ref{subsec:exp_setup}. Each generated simulation is evaluated according to the criteria defined in Table~\ref{tab:score_def}. The evaluation measures three aspects of system performance:
(1) \emph{cross-modal coherence}, measuring alignment among semantic inputs, generated motion, acoustic outputs, and environmental context;
(2) \emph{affective consistency}, measuring whether simulated behavioral and acoustic characteristics correspond to the affective descriptors specified in the input; and 
(3) \emph{simulation consistency}, measuring the stability of generated behaviors across repeated executions under identical prompts.

\subsubsection{Evaluation Datasets and Procedures}

18 prompts are evaluated, including 12 text-based prompts and 6 image-based prompts. Each prompt is executed five times, resulting in 90 generated simulations. The user-provided input consists only of the text or image prompt, while species selection, behavioral interpretation, and simulation parameters are determined by the LLM-based reasoning agents and transformed into UPR parameters. Generated simulations are manually evaluated according to the criteria defined in Table~\ref{tab:score_def}. Detailed per-prompt evaluation results are provided for text-based inputs in Table~\ref{tab:text_scored} and image-based inputs in Table~\ref{tab:image_scored}.

\subsubsection{Quantitative Results}

Table~\ref{tab:quant_eval_results_detail} summarizes the quantitative evaluation results. Repeated executions evaluate semantic alignment and simulation consistency under stochastic generation. The cross-modal coherence scores demonstrate the effectiveness of the UPR in maintaining consistency between semantic descriptions, simulated trajectories, vocalization patterns, and environmental context.

For text prompts (12 prompts, 60 runs), BirdsongChat achieves perfect cross-modal coherence and affective consistency, with average scores of 3.00 (100\%) for both metrics. Simulation consistency reaches 2.83 (94.4\%), indicating stable behavior generation while preserving controlled stochastic variations.

For image prompts (6 prompts, 30 runs), the system maintains perfect affective consistency with a score of 3.00 (100\%). Cross-modal coherence achieves 2.50 (83.3\%), with lower scores mainly resulting from ambiguity in visual interpretation, including environmental understanding and species selection. Simulation consistency remains high at 2.67 (88.9\%), demonstrating reliable repeated simulation under visually conditioned scenarios.

Across all evaluations (18 prompts, 90 runs), BirdsongChat achieves 2.83 (94.4\%) for cross-modal coherence, 3.00 (100\%) for affective consistency, and 2.78 (92.6\%) for simulation consistency. These results demonstrate that the UPR effectively maintains semantic, kinematic, acoustic, and environmental alignment while supporting stable multimodal simulation across repeated executions.

\subsubsection{Analysis and Discussion}

The quantitative results demonstrate the feasibility of the proposed hybrid multi-agent framework for connecting multimodal reasoning with physically based simulation. The UPR provides an explicit interface between reasoning and execution, encoding behavioral states and interpretable control parameters for coordinating motion, acoustics, and environmental rendering. As a prototype implementation, BirdsongChat demonstrates this design in interactive avian behavior simulation.

Evaluation across text and image inputs shows consistent affective alignment, with 100\% affective consistency for both modalities. Text inputs achieve higher cross-modal coherence (100\%) and generation consistency (94.4\%) than image inputs (83.3\% and 88.9\%), respectively. The lower image-based coherence suggests that visually complex or ambiguous environmental contexts remain challenging for semantic-to-physical grounding, while the 90.7\% overall image consistency indicates that UPR remains effective for heterogeneous inputs.

These results suggest that separating semantic interpretation from physical execution through an explicit intermediate representation can support interpretable and controllable behavior generation. Although demonstrated in an avian simulation setting, the proposed design provides a potential basis for embodied systems requiring coordinated translation of multimodal intent into physical behavior.

\section{Conclusion}

We presented a hybrid multi-agent framework for interactive cross-modal semantic-to-physical coordination in embodied environments. The framework uses a UPR to provide an interpretable interface between multimodal reasoning agents and physically based simulation agents. BirdsongChat, a prototype implementation of this framework, demonstrates its feasibility in interactive avian behavior simulation, demonstrating how shared behavioral and control parameters can coordinate motion, acoustics, and environmental rendering.

Across the evaluated text- and image-guided scenarios, the system achieves 94.4\% cross-modal coherence, 100\% affective consistency, and 92.6\% generation consistency. The results suggest that explicit intermediate representations can help preserve semantic intent across heterogeneous modalities while enabling controllable physical execution. More broadly, the proposed framework offers a general design principle for embodied AI systems that require interpretable coordination between multimodal reasoning and physical behavior.

\section{Limitations and Future Work}

Although the proposed framework provides an interpretable approach to semantic-to-physical simulation, several challenges remain. 

First, the reasoning layer depends on foundation models and is therefore susceptible to semantic ambiguity and reasoning errors that may propagate to downstream simulation. Future work will explore uncertainty-aware reasoning, verification mechanisms, and more robust multi-agent coordination.

Second, mapping high-level semantic descriptions to physical parameters remains challenging. While UPR provides an explicit interface, more adaptive representations are needed to generalize across species, behaviors, and application domains.

Third, the current evaluation is limited in scale. Our experiments provide an initial validation using 18 prompts with repeated executions and manual assessment based on predefined semantic, affective, and cross-modal consistency criteria. Larger evaluations with more diverse scenarios and independent human assessments are needed to examine behavioral plausibility, realism, and user perception.

Finally, BirdsongChat currently emphasizes controllable and perceptually plausible simulation rather than comprehensive ecological modeling. Future work should investigate richer ecological interactions, larger multi-agent environments, and real-time execution.

\section{Broader Impact and Ethics}

The proposed framework contributes to embodied AI research by exploring an interpretable interface between multimodal semantic reasoning and controllable physical simulation through coordinated reasoning and simulation agents. BirdsongChat, as a prototype implementation, demonstrates this approach in interactive avian behavior simulation. The framework may support future applications in bio-inspired ecoacoustics, swarm robotics, interactive simulation, virtual environments, ecological education, and creative media.

The generated bird behaviors, motion trajectories, and spatialized soundscapes are procedural simulations and should not be interpreted as faithful representations of real animals or natural ecosystems. When used in educational, ecological, or media contexts, such content should be clearly identified as synthetic. In addition, foundation-model-based reasoning agents may introduce semantic ambiguity, biases, or unintended interpretations that can propagate to physical simulation. Responsible deployment should therefore emphasize transparent agent coordination, interpretable semantic-to-physical mappings, validation of generated behaviors, and clear communication of the synthetic nature and limitations of the resulting content.

\bibliography{custom}

@article{10.1145/3717059,
author = {Liu, Huaping and Guo, Di and Cangelosi, Angelo},
title = {Embodied Intelligence: A Synergy of Morphology, Action, Perception and Learning},
year = {2025},
issue_date = {July 2025},
publisher = {Association for Computing Machinery},
address = {New York, NY, USA},
volume = {57},
number = {7},
issn = {0360-0300},
url = {https://doi.org/10.1145/3717059},
doi = {10.1145/3717059},
journal = {ACM Comput. Surv.},
month = mar,
articleno = {186},
numpages = {36}
}

@inproceedings{liao-etal-2025-agentmaster,
    title = "{A}gent{M}aster: A Multi-Agent Conversational Framework Using {A}2{A} and {MCP} Protocols for Multimodal Information Retrieval and Analysis",
    author = "Liao, Callie C.  and
      Liao, Duoduo  and
      Gadiraju, Sai Surya",
    editor = {Habernal, Ivan  and
      Schulam, Peter  and
      Tiedemann, J{\"o}rg},
    booktitle = "Proceedings of the 2025 Conference on Empirical Methods in Natural Language Processing: System Demonstrations",
    month = nov,
    year = "2025",
    address = "Suzhou, China",
    publisher = "Association for Computational Linguistics",
    url = "https://aclanthology.org/2025.emnlp-demos.5/",
    doi = "10.18653/v1/2025.emnlp-demos.5",
    pages = "52--72",
    ISBN = "979-8-89176-334-0",
}

@INPROCEEDINGS{11402190,
  author={Zhang, Ellie L. and Liao, Duoduo and Liao, Callie C.},
  booktitle={2025 IEEE International Conference on Big Data (BigData)}, 
  title={Dynamic Multi-Species Bird Soundscape Generation with Acoustic Patterning and 3{D} Spatialization}, 
  year={2025},
  volume={},
  number={},
  pages={5057-5066},
  doi={10.1109/BigData66926.2025.11402190}}

@book{oppenheim2010discrete,
  title={Discrete-Time Signal Processing},
  author={Oppenheim, Alan V and Schafer, Ronald W},
  year={2010},
  edition={3rd},
  publisher={Pearson}
}

@InProceedings{pmlr-v270-kim25c,
  title = 	 {Open{VLA}: An Open-Source Vision-Language-Action Model},
  author =       {Kim, Moo Jin and Pertsch, Karl and Karamcheti, Siddharth and Xiao, Ted and Balakrishna, Ashwin and Nair, Suraj and Rafailov, Rafael and Foster, Ethan P and Sanketi, Pannag R and Vuong, Quan and Kollar, Thomas and Burchfiel, Benjamin and Tedrake, Russ and Sadigh, Dorsa and Levine, Sergey and Liang, Percy and Finn, Chelsea},
  booktitle = 	 {Proceedings of The 8th Conference on Robot Learning},
  pages = 	 {2679--2713},
  year = 	 {2025},
  editor = 	 {Agrawal, Pulkit and Kroemer, Oliver and Burgard, Wolfram},
  volume = 	 {270},
  series = 	 {Proceedings of Machine Learning Research},
  month = 	 {06--09 Nov},
  publisher =    {PMLR},
  url = 	 {https://proceedings.mlr.press/v270/kim25c.html},
}

@inproceedings{wang-etal-2025-mio,
    title = "{MIO}: A Foundation Model on Multimodal Tokens",
    author = "Wang, Zekun Moore  and
      Zhu, King  and
      Xu, Chunpu  and
      Zhou, Wangchunshu  and
      Liu, Jiaheng  and
      Zhang, Yibo  and
      Wang, Jessie  and
      Shi, Ning  and
      Li, Siyu  and
      Li, Yizhi  and
      Que, Haoran  and
      Zhang, Zhaoxiang  and
      Zhang, Yuanxing  and
      Zhang, Ge  and
      Xu, Ke  and
      Fu, Jie  and
      Huang, Wenhao",
    editor = "Christodoulopoulos, Christos  and
      Chakraborty, Tanmoy  and
      Rose, Carolyn  and
      Peng, Violet",
    booktitle = "Proceedings of the 2025 Conference on Empirical Methods in Natural Language Processing",
    month = nov,
    year = "2025",
    address = "Suzhou, China",
    publisher = "Association for Computational Linguistics",
    url = "https://aclanthology.org/2025.emnlp-main.255/",
    doi = "10.18653/v1/2025.emnlp-main.255",
    pages = "5077--5099",
}

@INPROCEEDINGS{11094075,
  author={Li, Shufan and Kallidromitis, Konstantinos and Gokul, Akash and Liao, Zichun and Kato, Yusuke and Kozuka, Kazuki and Grover, Aditya},
  booktitle={2025 IEEE/CVF Conference on Computer Vision and Pattern Recognition (CVPR)}, 
  title={OmniFlow: Any-to-Any Generation with Multi-Modal Rectified Flows}, 
  year={2025},
  volume={},
  number={},
  pages={13178-13188},
  doi={10.1109/CVPR52734.2025.01230}}

@inproceedings{ICLR2025_e9e140df,
 author = {Wu, Yecheng and Zhang, Zhuoyang and Chen, Junyu and Tang, Haotian and Li, Dacheng and Fang, Yunhao and Zhu, Ligeng and Xie, Enze and Yin, Hongxu and Yi, Li and Han, Song and Lu, Yao},
 booktitle = {International Conference on Learning Representations},
 editor = {Y. Yue and A. Garg and N. Peng and F. Sha and R. Yu},
 pages = {93620--93638},
 title = {{VILA-U}: a Unified Foundation Model Integrating Visual Understanding and Generation},
 url = {https://proceedings.iclr.cc/paper_files/paper/2025/file/e9e140df6de01afb672cb859d203c307-Paper-Conference.pdf},
 volume = {2025},
 year = {2025}
}

@inproceedings{NEURIPS2022_960a172b,
 author = {Alayrac, Jean-Baptiste and Donahue, Jeff and Luc, Pauline and Miech, Antoine and Barr, Iain and Hasson, Yana and Lenc, Karel and Mensch, Arthur and Millican, Katherine and Reynolds, Malcolm and Ring, Roman and Rutherford, Eliza and Cabi, Serkan and Han, Tengda and Gong, Zhitao and Samangooei, Sina and Monteiro, Marianne and Menick, Jacob L and Borgeaud, Sebastian and Brock, Andy and Nematzadeh, Aida and Sharifzadeh, Sahand and Bi\'{n}kowski, Miko\l aj and Barreira, Ricardo and Vinyals, Oriol and Zisserman, Andrew and Simonyan, Kar\'{e}n},
 booktitle = {Advances in Neural Information Processing Systems},
 doi = {10.52202/068431-1723},
 editor = {S. Koyejo and S. Mohamed and A. Agarwal and D. Belgrave and K. Cho and A. Oh},
 pages = {23716--23736},
 publisher = {Curran Associates, Inc.},
 title = {Flamingo: a Visual Language Model for Few-Shot Learning},
 url = {https://proceedings.neurips.cc/paper_files/paper/2022/file/960a172bc7fbf0177ccccbb411a7d800-Paper-Conference.pdf},
 volume = {35},
 year = {2022}
}

@InProceedings{pmlr-v202-li23q,
  title = 	 {{BLIP}-2: Bootstrapping Language-Image Pre-training with Frozen Image Encoders and Large Language Models},
  author =       {Li, Junnan and Li, Dongxu and Savarese, Silvio and Hoi, Steven},
  booktitle = 	 {Proceedings of the 40th International Conference on Machine Learning},
  pages = 	 {19730--19742},
  year = 	 {2023},
  editor = 	 {Krause, Andreas and Brunskill, Emma and Cho, Kyunghyun and Engelhardt, Barbara and Sabato, Sivan and Scarlett, Jonathan},
  volume = 	 {202},
  series = 	 {Proceedings of Machine Learning Research},
  month = 	 {23--29 Jul},
  publisher =    {PMLR},
  url = 	 {https://proceedings.mlr.press/v202/li23q.html},

}

@InProceedings{pmlr-v205-ichter23a,
  title = 	 {{Do As I Can, Not As I Say}: Grounding Language in Robotic Affordances},
  author =       {Ichter, Brian and Brohan, Anthony and Chebotar, Yevgen and Finn, Chelsea and Hausman, Karol and Herzog, Alexander and Ho, Daniel and Ibarz, Julian and Irpan, Alex and Jang, Eric and Julian, Ryan and Kalashnikov, Dmitry and Levine, Sergey and Lu, Yao and Parada, Carolina and Rao, Kanishka and Sermanet, Pierre and Toshev, Alexander T and Vanhoucke, Vincent and Xia, Fei and Xiao, Ted and Xu, Peng and Yan, Mengyuan and Brown, Noah and Ahn, Michael and Cortes, Omar and Sievers, Nicolas and Tan, Clayton and Xu, Sichun and Reyes, Diego and Rettinghouse, Jarek and Quiambao, Jornell and Pastor, Peter and Luu, Linda and Lee, Kuang-Huei and Kuang, Yuheng and Jesmonth, Sally and Joshi, Nikhil J. and Jeffrey, Kyle and Ruano, Rosario Jauregui and Hsu, Jasmine and Gopalakrishnan, Keerthana and David, Byron and Zeng, Andy and Fu, Chuyuan Kelly},
  booktitle = 	 {Proceedings of The 6th Conference on Robot Learning},
  pages = 	 {287--318},
  year = 	 {2023},
  editor = 	 {Liu, Karen and Kulic, Dana and Ichnowski, Jeff},
  volume = 	 {205},
  series = 	 {Proceedings of Machine Learning Research},
  month = 	 {14--18 Dec},
  publisher =    {PMLR},
  url = 	 {https://proceedings.mlr.press/v205/ichter23a.html},
}

@InProceedings{pmlr-v229-zitkovich23a,
  title = 	 {{RT}-2: Vision-Language-Action Models Transfer Web Knowledge to Robotic Control},
  author =       {Zitkovich, Brianna and Yu, Tianhe and Xu, Sichun and Xu, Peng and Xiao, Ted and Xia, Fei and Wu, Jialin and Wohlhart, Paul and Welker, Stefan and Wahid, Ayzaan and Vuong, Quan and Vanhoucke, Vincent and Tran, Huong and Soricut, Radu and Singh, Anikait and Singh, Jaspiar and Sermanet, Pierre and Sanketi, Pannag R. and Salazar, Grecia and Ryoo, Michael S. and Reymann, Krista and Rao, Kanishka and Pertsch, Karl and Mordatch, Igor and Michalewski, Henryk and Lu, Yao and Levine, Sergey and Lee, Lisa and Lee, Tsang-Wei Edward and Leal, Isabel and Kuang, Yuheng and Kalashnikov, Dmitry and Julian, Ryan and Joshi, Nikhil J. and Irpan, Alex and Ichter, Brian and Hsu, Jasmine and Herzog, Alexander and Hausman, Karol and Gopalakrishnan, Keerthana and Fu, Chuyuan and Florence, Pete and Finn, Chelsea and Dubey, Kumar Avinava and Driess, Danny and Ding, Tianli and Choromanski, Krzysztof Marcin and Chen, Xi and Chebotar, Yevgen and Carbajal, Justice and Brown, Noah and Brohan, Anthony and Arenas, Montserrat Gonzalez and Han, Kehang},
  booktitle = 	 {Proceedings of The 7th Conference on Robot Learning},
  pages = 	 {2165--2183},
  year = 	 {2023},
  editor = 	 {Tan, Jie and Toussaint, Marc and Darvish, Kourosh},
  volume = 	 {229},
  series = 	 {Proceedings of Machine Learning Research},
  month = 	 {06--09 Nov},
  publisher =    {PMLR},
  url = 	 {https://proceedings.mlr.press/v229/zitkovich23a.html},
}

@misc{yao2023reactsynergizingreasoningacting,
      title={Re{A}ct: Synergizing Reasoning and Acting in Language Models}, 
      author={Shunyu Yao and Jeffrey Zhao and Dian Yu and Nan Du and Izhak Shafran and Karthik Narasimhan and Yuan Cao},
      year={2023},
      eprint={2210.03629},
      archivePrefix={arXiv},
      primaryClass={cs.CL},
      url={https://arxiv.org/abs/2210.03629}, 
}

@misc{wang2023voyageropenendedembodiedagent,
      title={Voyager: An Open-Ended Embodied Agent with Large Language Models}, 
      author={Guanzhi Wang and Yuqi Xie and Yunfan Jiang and Ajay Mandlekar and Chaowei Xiao and Yuke Zhu and Linxi Fan and Anima Anandkumar},
      year={2023},
      eprint={2305.16291},
      archivePrefix={arXiv},
      primaryClass={cs.AI},
      url={https://arxiv.org/abs/2305.16291}, 
}

@inproceedings{10.1145/3586183.3606763,
author = {Park, Joon Sung and O'Brien, Joseph and Cai, Carrie Jun and Morris, Meredith Ringel and Liang, Percy and Bernstein, Michael S.},
title = {Generative {A}gents: Interactive Simulacra of Human Behavior},
year = {2023},
isbn = {9798400701320},
publisher = {Association for Computing Machinery},
address = {New York, NY, USA},
url = {https://doi.org/10.1145/3586183.3606763},
doi = {10.1145/3586183.3606763},
booktitle = {Proceedings of the 36th Annual ACM Symposium on User Interface Software and Technology},
articleno = {2},
numpages = {22},
location = {San Francisco, CA, USA},
series = {UIST '23}
}

@article{UhlenbeckOrnstein30,
  author    = {Uhlenbeck , George E. and Ornstein, Leonard S. },
  title     = {On the Theory of the Brownian Motion},
  journal   = {Physical Review},
  volume    = {36},
  number    = {5},
  pages     = {823--841},
  year      = {1930},
  doi       = {10.1103/PhysRev.36.823},
  url = {https://journals.aps.org/pr/abstract/10.1103/PhysRev.36.823}
}

@book{KloedenPlaten92,
  author    = {Kloeden, Peter E.  and Platen, Eckhard },
  title     = {Numerical Solution of Stochastic Differential Equations},
  publisher = {Springer},
  year      = {1992},
  address   = {Berlin},
  url = {https://link.springer.com/book/10.1007/978-3-662-12616-5},
  edition   = {1st}
}

@InProceedings{pmlr-v202-driess23a,
  title = 	 {{P}a{LM}-{E}: An Embodied Multimodal Language Model},
  author =       {Driess, Danny and Xia, Fei and Sajjadi, Mehdi S. M. and Lynch, Corey and Chowdhery, Aakanksha and Ichter, Brian and Wahid, Ayzaan and Tompson, Jonathan and Vuong, Quan and Yu, Tianhe and Huang, Wenlong and Chebotar, Yevgen and Sermanet, Pierre and Duckworth, Daniel and Levine, Sergey and Vanhoucke, Vincent and Hausman, Karol and Toussaint, Marc and Greff, Klaus and Zeng, Andy and Mordatch, Igor and Florence, Pete},
  booktitle = 	 {Proceedings of the 40th International Conference on Machine Learning},
  pages = 	 {8469--8488},
  year = 	 {2023},
  editor = 	 {Krause, Andreas and Brunskill, Emma and Cho, Kyunghyun and Engelhardt, Barbara and Sabato, Sivan and Scarlett, Jonathan},
  volume = 	 {202},
  series = 	 {Proceedings of Machine Learning Research},
  month = 	 {23--29 Jul},
  publisher =    {PMLR},
  url = 	 {https://proceedings.mlr.press/v202/driess23a.html},

}

@INPROCEEDINGS{10943341,
  author={Fu, Rao and Liu, Jingyu and Chen, Xilun and Nie, Yixin and Xiong, Wenhan},
  booktitle={2025 IEEE/CVF Winter Conference on Applications of Computer Vision (WACV)}, 
  title={Scene-{LLM}: Extending Language Model for 3D Visual Reasoning}, 
  year={2025},
  volume={},
  number={},
  pages={2195-2206},
  doi={10.1109/WACV61041.2025.00220}}

@inproceedings{NEURIPS2023_94b472a1,
 author = {Copet, Jade and Kreuk, Felix and Gat, Itai and Remez, Tal and Kant, David and Synnaeve, Gabriel and Adi, Yossi and Defossez, Alexandre},
 booktitle = {Advances in Neural Information Processing Systems},
 editor = {A. Oh and T. Naumann and A. Globerson and K. Saenko and M. Hardt and S. Levine},
 pages = {47704--47720},
 publisher = {Curran Associates, Inc.},
 title = {Simple and Controllable Music Generation},
 url = {https://proceedings.neurips.cc/paper_files/paper/2023/file/94b472a1842cd7c56dcb125fb2765fbd-Paper-Conference.pdf},
 volume = {36},
 year = {2023}
}

@misc{agostinelli2023musiclmgeneratingmusictext,
      title={Music{LM}: Generating Music From Text}, 
      author={Andrea Agostinelli and Timo I. Denk and Zalán Borsos and Jesse Engel and Mauro Verzetti and Antoine Caillon and Qingqing Huang and Aren Jansen and Adam Roberts and Marco Tagliasacchi and Matt Sharifi and Neil Zeghidour and Christian Frank},
      year={2023},
      eprint={2301.11325},
      archivePrefix={arXiv},
      primaryClass={cs.SD},
      url={https://arxiv.org/abs/2301.11325}, 
}

@inproceedings{10.1609/aaai.v39i24.34750,
author = {Yang, Chenyu and Wang, Shuai and Chen, Hangting and Yu, Jianwei and Tan, Wei and Gu, Rongzhi and Xu, Yaoxun and Zhou, Yizhi and Zhu, Haina and Li, Haizhou},
title = {Song{Ed}itor: adapting zero-shot song generation language model as a multi-task editor},
year = {2025},
isbn = {978-1-57735-897-8},
publisher = {AAAI Press},
url = {https://doi.org/10.1609/aaai.v39i24.34750},
doi = {10.1609/aaai.v39i24.34750},
booktitle = {Proceedings of the Thirty-Ninth AAAI Conference on Artificial Intelligence and Thirty-Seventh Conference on Innovative Applications of Artificial Intelligence and Fifteenth Symposium on Educational Advances in Artificial Intelligence},
articleno = {2852},
numpages = {9},
series = {AAAI'25/IAAI'25/EAAI'25}
}

@inproceedings{NEURIPS2024_cebbd24f,
 author = {Huang, Haifeng and Chen, Yilun and Wang, Zehan and Huang, Rongjie and Xu, Runsen and Wang, Tai and Liu, Luping and Cheng, Xize and Zhao, Yang and Pang, Jiangmiao and Zhao, Zhou},
 booktitle = {Advances in Neural Information Processing Systems},
 doi = {10.52202/079017-3620},
 editor = {A. Globerson and L. Mackey and D. Belgrave and A. Fan and U. Paquet and J. Tomczak and C. Zhang},
 pages = {113991--114017},
 publisher = {Curran Associates, Inc.},
 title = {Chat-{S}cene: Bridging 3{D} Scene and Large Language Models with Object Identifiers},
 url = {https://proceedings.neurips.cc/paper_files/paper/2024/file/cebbd24f1e50bcb63d015611fe0fe767-Paper-Conference.pdf},
 volume = {37},
 year = {2024}
}

@article{10.1145/3658146,
author = {Zhang, Longwen and Wang, Ziyu and Zhang, Qixuan and Qiu, Qiwei and Pang, Anqi and Jiang, Haoran and Yang, Wei and Xu, Lan and Yu, Jingyi},
title = {{CLAY}: A Controllable Large-scale Generative Model for Creating High-quality 3D Assets},
year = {2024},
issue_date = {July 2024},
publisher = {Association for Computing Machinery},
address = {New York, NY, USA},
volume = {43},
number = {4},
issn = {0730-0301},
url = {https://doi.org/10.1145/3658146},
doi = {10.1145/3658146},
journal = {ACM Trans. Graph.},
month = jul,
articleno = {120},
numpages = {20}
}

@inproceedings{10.1145/37401.37406,
author = {Reynolds, Craig W.},
title = {Flocks, herds and schools: A distributed behavioral model},
year = {1987},
isbn = {0897912276},
publisher = {Association for Computing Machinery},
address = {New York, NY, USA},
url = {https://doi.org/10.1145/37401.37406},
doi = {10.1145/37401.37406},
booktitle = {Proceedings of the 14th Annual Conference on Computer Graphics and Interactive Techniques},
pages = {25–34},
numpages = {10},
series = {SIGGRAPH '87}
}
\bibliographystyle{IEEEtran}
% \newpage
% \clearpage
 % \appendix

% \onecolumn
\newpage

\end{document}